\documentclass{PoS}

\usepackage{amsmath}

\title{
\vspace{-3cm}
\hspace*{\fill}{\normalsize \tt MPP-2010-3}\\
\vspace{3cm}
Radiative corrections to $W\gamma\gamma$ production at the LHC}

\ShortTitle{Radiative corrections to $W\gamma\gamma$ production at the LHC}

\author{Uli Baur, Doreen Wackeroth\\
  Department of Physics, State University of New York, Buffalo, NY 14260, USA\\
  E-mail: \email{baur@ubhep.physics.buffalo.edu},\\
  \phantom{E-mail:} \email{dow@ubpheno.physics.buffalo.edu}}

\author{\speaker{Marcus M. Weber}\\
 Max-Planck-Institut f\"ur Physik, F\"ohringer Ring 6, 80805 Munich, Germany\\
E-mail: \email{mmweber@mppmu.mpg.de}}

\abstract{
Radiative W production at hadron colliders is an important testing
ground for the Standard Model. We consider $W\gamma\gamma$ production
which is sensitive to the quartic $WW\gamma\gamma$ coupling.
Furthermore the Standard Model amplitude for this process contains a radiation
zero. We present a calculation of the NLO QCD corrections for
$W\gamma\gamma$ production at the LHC.}

\FullConference{RADCOR 2009 - 9th International Symposium on Radiative Corrections (Applications of Quantum Field Theory to Phenomenology) \\
		 October 25-30 2009\\
		 Ascona, Switzerland}

\newcommand{\gev}{\ensuremath{\,\text{GeV}}}

\newcommand{\tev}{\ensuremath{\,\text{TeV}}}

\newcommand{\fb}{\ensuremath{\,\text{fb}}}

\newcommand{\ppwyy}{\ensuremath{pp \to W\gamma\gamma}}
\newcommand{\ppwpyy}{\ensuremath{pp \to W^+\gamma\gamma}}
\newcommand{\wy}{\ensuremath{W\gamma}}
\newcommand{\wyy}{\ensuremath{W\gamma\gamma}}
\newcommand{\wpyy}{\ensuremath{W^+\gamma\gamma}}
\newcommand{\wwyy}{\ensuremath{WW\gamma\gamma}}

\newcommand{\FA}{\textsc{FeynArts}{}}
\newcommand{\FC}{\textsc{FormCalc}{}}
\newcommand{\madgraph}{\textsc{MadGraph}{}}
\newcommand{\qd}{\textsc{QD}{}}

\newcommand{\mref}[1]{Ref.~\cite{#1}}
\newcommand{\tabre}[1]{Tab.~\ref{#1}}
\newcommand{\figre}[1]{Fig.~\ref{#1}}

\begin{document}


\section{Introduction}

With the startup of the LHC the mechanism of electroweak symmetry breaking
is accessible to direct experimental investigation.  In the Standard Model
electroweak symmetry breaking is realized by the Higgs mechanism leading to
a single scalar physical state, the Higgs boson. If this description is
correct the LHC will be able to discover the Higgs boson and measure some
of its properties.  It is also possible that the LHC will uncover evidence of
physics beyond the Standard Model either by the direct production of new
states or by deviations of the couplings of known particles from their
Standard Model values.

Anomalous interactions of the known gauge bosons are one interesting
possibility and can be studied in gauge-boson production
processes. While double gauge-boson production gives access to the
triple gauge couplings, triple gauge-boson production processes allow
to investigate quartic gauge couplings.  The QCD corrections to the
production of three heavy gauge bosons have been calculated
\cite{Lazopoulos:2007ix}, in some cases including the leptonic decays, and
turn out to be fairly large.  The corrections to $WW\gamma$ and $ZZ\gamma$ production
at the LHC have also become available recently \cite{Bozzi:2009ig} and are
sizeable.

Another interesting process is $\wyy$ production which is sensitive to
both the triple $\wyy$ and the quartic $\wwyy$ coupling
\cite{Baur:1997bn, Bell:2009vh}.  As already the case
for $\wy$ production this process exhibits a radiation zero, i.e. the
leading order amplitude vanishes exactly for some momentum
configurations. The radiation zero in the amplitude appears at
a specific polar angle of the $W$ with respect to the incoming quark
direction in the partonic centre-of-mass frame if the photons
are collinear. This zero will be
washed out in practice by the the difficulty in reconstructing the
partonic c.m. frame, the unknown incoming quark direction
and finite detector resolution effects.  Furthermore it could be
filled by radiative corrections.

In view of the large radiative corrections for similar processes a full NLO
QCD calculation is needed for reliable theoretical predictions. We present
the QCD corrections to $\wyy$ production at the LHC treating the $W$ as
stable, i.e. without including
decays of the $W$.  Our calculation has been implemented in a flexible
Monte Carlo computer code allowing to calculate arbitrary distributions
from a set of weighted events.



\section{Calculation}

Since we are interested in the anomalous quartic gauge coupling and
the radiation zero we focus on the production of two isolated photons
accompanying the $W$.  Producing both photons in the hard interaction
gives rise to the direct contribution to $\ppwyy$. The same final
state can also be reached by producing $Wq\gamma$ in the hard
interaction followed by a non-perturbative splitting of the quark into
a photon. This is described by a fragmentation function
$D_{\gamma/q}(z)$ which is formally of ${\cal
  O}(\frac{\alpha}{\alpha_s})$. Both the direct and the fragmentation
contribution are therefore formally of the same order in the
couplings. In fact only the sum of direct and fragmentation
contributions is physically meaningful.  In the fragmentation
contribution the photon will be accompanied by a collinear hadronic
remnant. This contribution can therefore be suppressed by requiring
the photons to be isolated which will not affect the direct
contribution.

At NLO both the corrections to the direct and the fragmentation
contributions are needed in principle. The fragmentation contribution
will produce a fragmentation counterterm which cancels the QED
singularity from the $q \to q\gamma$ splitting in the real radiation
process $qg \to Wq\gamma$. The QED singularity is therefore absorbed
into a redefinition of the fragmentation function at NLO.  Due to the
effective suppression of the fragmentation contribution using photon
isolation cuts the NLO corrections to the fragmentation process
contribute only little. We therefore take into account only the NLO
corrections to the direct contribution.  The fragmentation
contribution is calculated only at LO but with the NLO fragmentation
functions of \mref{Duke:1982bj}, i.e. the fragmentation counterterm is included.

The calculation itself has been performed in a diagrammatic approach
mostly using standard techniques and tools.  For the virtual
corrections the generation and simplification of the diagrams has been
performed using \FA\ \cite{Hahn:2000kx} and \FC\ \cite{Hahn:1998yk}.
The analytical results of \FC\ have then been
translated to C++ code. For the real amplitudes \madgraph\ has been
used \cite{Stelzer:1994ta}. In order to extract the soft and collinear
singularities from the real corrections and combine them with the
virtual contribution we have implemented both a two cutoff phase-space
slicing method according to \cite{Harris:2001sx} and the dipole
subtraction formalism \cite{Catani:1996vz}.

The tensor 1-loop integrals appearing in the virtual amplitudes are
evaluated using a tensor reduction approach. The 5-point integrals are
reduced to scalar 4-point functions in a numerically stable way using
\mref{Denner:2002ii}. For the 3- and 4-point tensor integrals we
employ Passarino-Veltman reduction \cite{Passarino:1978jh}. In regions
of phase space where this becomes unstable we still use the same
reduction but switch to high-precision arithmetic using the
\qd\ library \cite{Bailey:qd}. Since only a small fraction of the points are
unstable the additional time needed by the high-precision evaluations
is moderate.



\section{Numerical Results}

\begin{table}
\centerline{
\begin{tabular}{l r@{.}l r@{.}l r@{.}l}
& \multicolumn{6}{c}{$\sigma(\ppwpyy)\,[\fb]$} \\
&  \multicolumn{2}{c}{standard} & \multicolumn{2}{c}{with isolation}  &  \multicolumn{2}{c}{iso \& jet veto} \\ \hline
LO direct &   7&253(5) &   7&253(5)  &  7&253(5) \\
LO frag   &  24&30(2)  &   1&505(1)  &  1&501(1) \\ \hline
LO total  &  31&55     &   8&758     &  8&754 \\ \hline
NLO	  &  39&33(6)  &  25&62(4)   & 11&83(4) \\ \hline
K factor  &   1&25     &   2&93      &  1&35
\end{tabular}}
\caption{Total cross sections for $\ppwpyy$ at LHC for $\sqrt{s} = 14\tev$ with standard cuts only, with additional photon isolation cuts and with photon isolation and jet veto cuts. Shown are the leading-order direct and fragmentation contributions, the total leading-order cross section, the complete NLO result and the K-factor.}
\label{tab}
\end{table}

We now turn to the study of some results for $\wpyy$ production at the
LHC for $\sqrt{s} = 14 \tev$.
To obtain a well-defined cross section  we impose the
following standard cuts
\[
p_{T,\gamma} > 30 \gev \qquad |\eta_\gamma| < 2.5 \qquad
\Delta R_{\gamma\gamma} > 0.4 \qquad \Delta R = \sqrt{\Delta \phi ^2 + \Delta \eta ^2}
\]
on the transverse momenta, the pseudorapidities and the separation of the photons.
The corresponding cross section and the associated K-factor are shown
in \tabre{tab}. For these cuts the fragmentation contribution
is by far dominant. The NLO corrections are only moderate when
compared to the combined LO result. However no corrections to the
fragmentation contribution have been taken into account in our
calculation.

In order to isolate the direct contribution we impose an additional
photon isolation criterion. This is implemented by requiring the
hadronic transverse energy inside a cone around the photon to be limited
\[
E_{T,had} < \epsilon E_{T,\gamma} \quad \text{inside cone} \quad \Delta R(\gamma,had) < R_{cone}.
\]
We use the energy fraction $\epsilon = 0.15$ and a cone size of
$R_\mathrm{cone} = 0.7$. As \tabre{tab} shows this suppresses the
fragmentation contribution by more than an order of magnitude while
leaving the direct contribution unchanged. The NLO corrections are
very large with a K-factor of 2.93. At NLO
quark-gluon and antiquark-gluon induced partonic channels open up in
the real corrections which might be responsible for the large
corrections.

In order to verify if the large corrections are caused by the real
corrections we impose a jet veto in addition to the standard and
isolation cuts. We remove all events with the jets fulfilling
\[
\quad p_{Tj} > 50 \gev \quad \text{and} \quad  |\eta_j| < 3.
\]
This leaves the leading order unchanged while greatly reducing the QCD
corrections as can be seen from \tabre{tab}. The corrections are now
only 35\% and the large corrections seen without jet veto are
therefore indeed caused by real radiation.

\begin{figure}[t]
\centerline{
\includegraphics[width=0.36\textwidth]{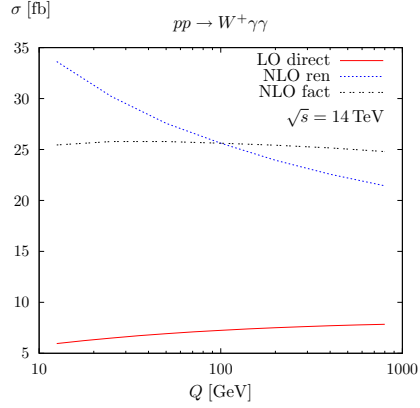}}
\caption{Scale dependence of the direct leading-order and the full NLO
  cross sections using the standard and photon isolation cuts. No jet veto
  is applied. The renormalization and factorization scale dependencies of
  the NLO result are shown separately.}
\label{fig:scaledep}
\end{figure}
The scale dependence at leading and next-to-leading order is shown in
\figre{fig:scaledep}. Since the LO amplitude does not contain $\alpha_s$
the renormalization scale dependence is increased at NLO. The factorization
scale dependence is stabilized however.

\begin{figure}[t]
\centerline{
  \includegraphics[width=0.39\textwidth]{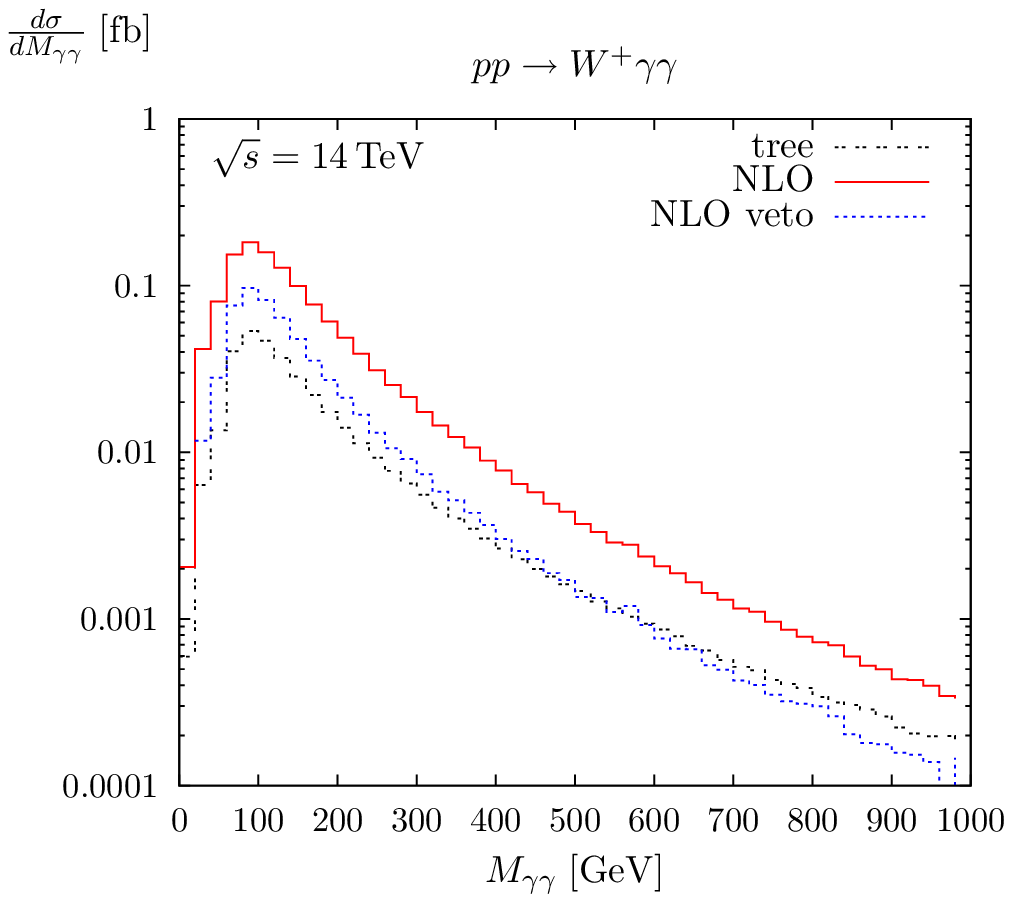}
\qquad
  \includegraphics[width=0.39\textwidth]{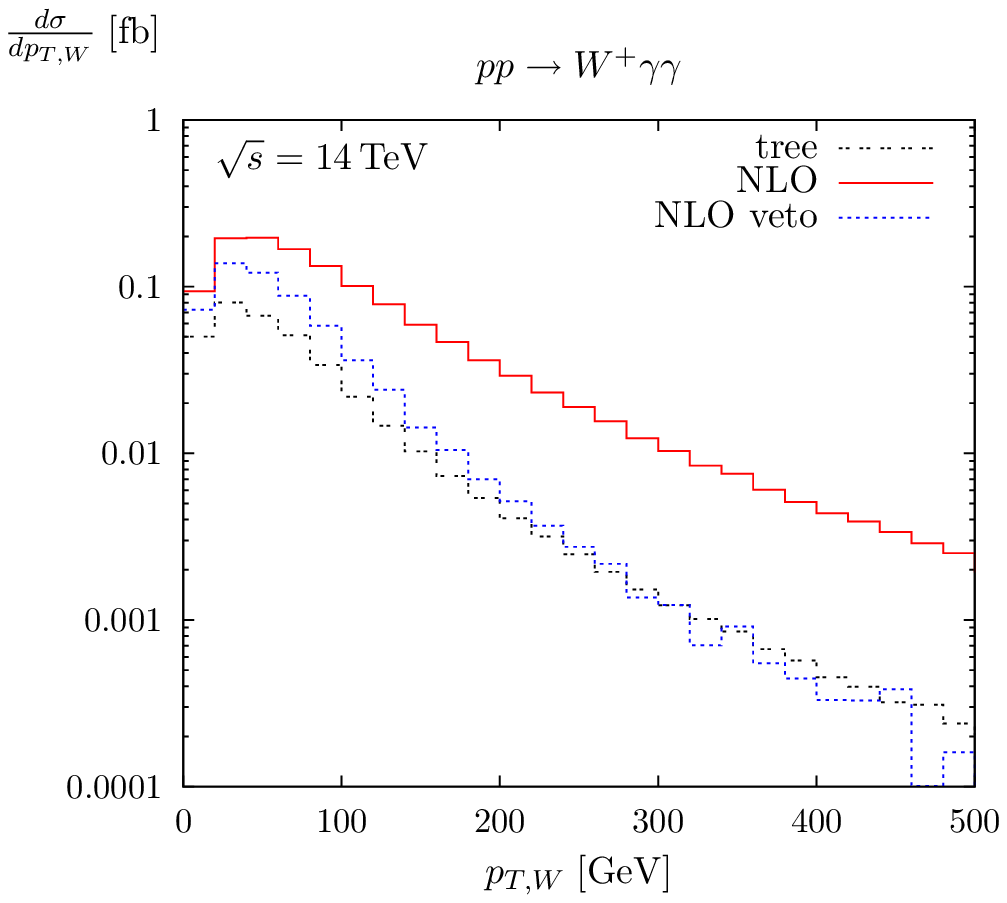}
}
\caption{Distributions in $\gamma\gamma$ invariant mass and $W$ transverse
  momentum at leading and next-to-leading order using standard and photon
  isolation cuts. Also shown is the NLO result using an additional jet veto.}
\label{fig:histo1}
\end{figure}
In \figre{fig:histo1} distributions in the invariant mass
$M_{\gamma\gamma}$ of the photons and the transverse $W$ momentum $p_{T,W}$
are shown. The corrections show a shape dependence in both cases. This is
especially strong in the $p_{T,W}$ distribution with a K-factor of about
2.5 at $p_T = 20 \gev$ and about 10 at $p_T = 500 \gev$. The effect of the
jet veto is to remove the positive corrections at high invariant masses and
$p_T$ completely while not affecting the lower regions as much. Regardless
of the use of a jet veto the NLO corrections show a strong shape dependence
and can therefore not be described by a global K-factor.

\begin{figure}[t]
\centerline{
  \includegraphics[width=0.37\textwidth]{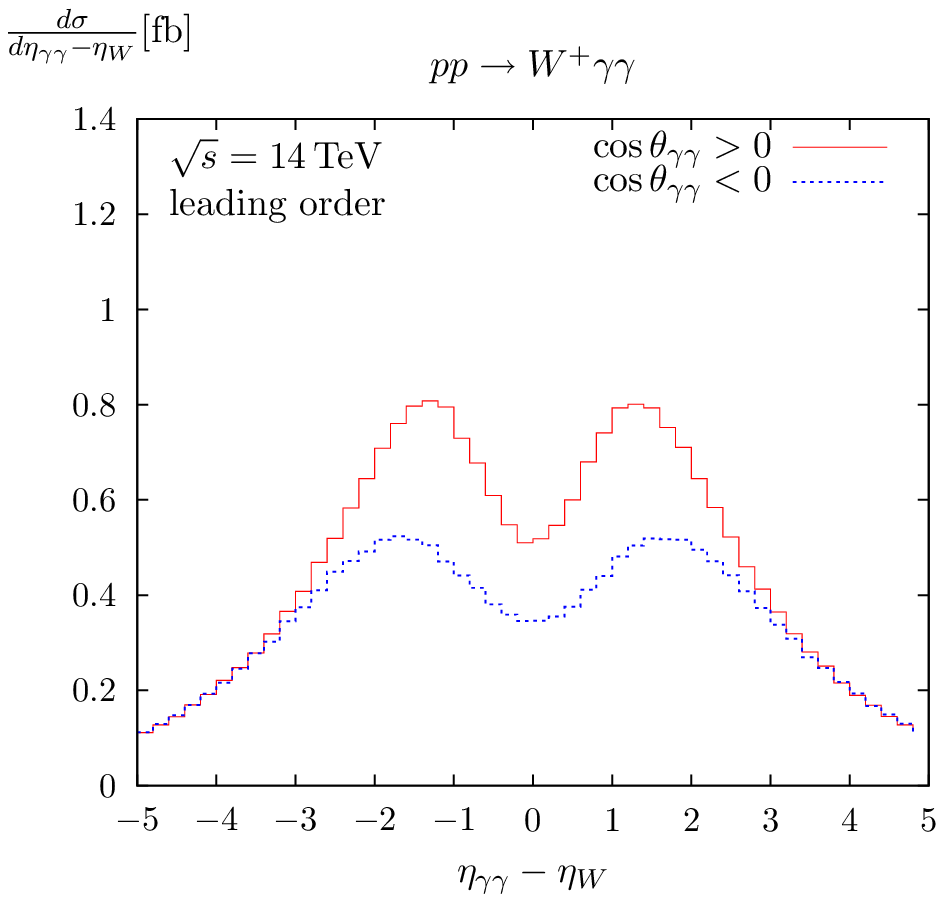}
\qquad
  \includegraphics[width=0.37\textwidth]{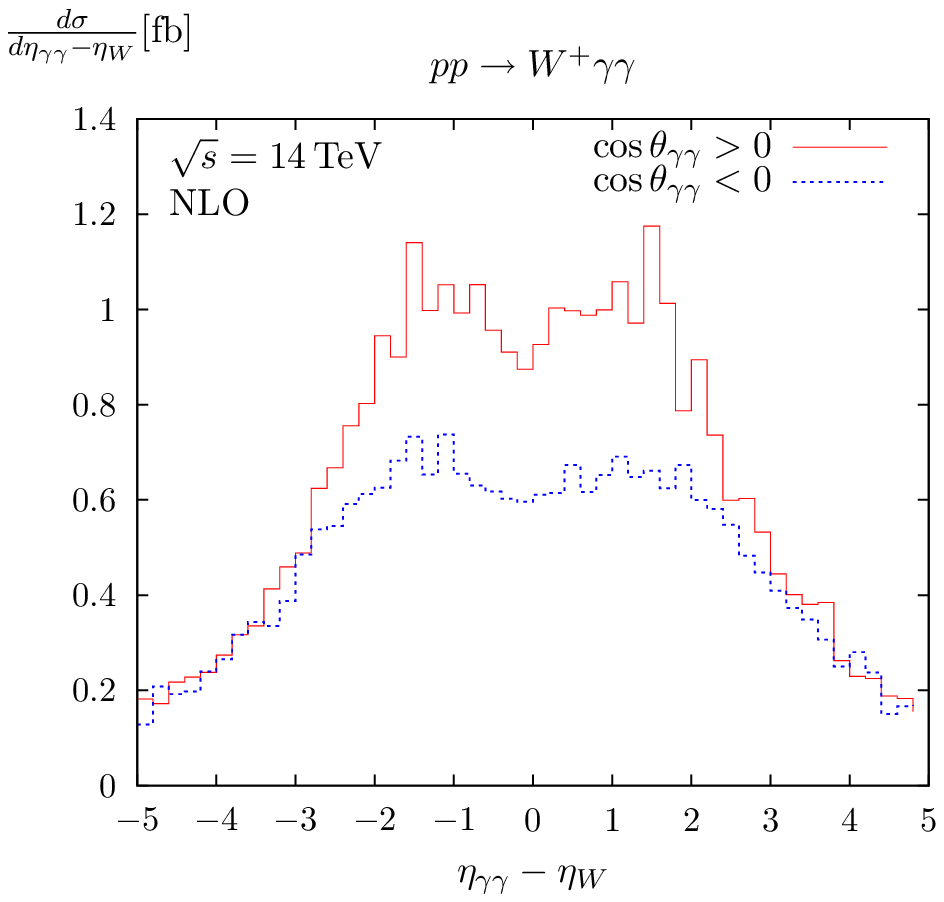}
}
\caption{Distributions in pseudorapidity separation between the $W$ and
  the two-photon system for the photons in the same and in opposite
  hemispheres at leading order (l.h.s) and NLO (r.h.s). Both
  photon isolation and jet veto cuts have been imposed.}
\label{fig:histo2}
\end{figure}
To investigate the impact of the NLO corrections on the radiation zero we
show the separation in pseudorapidity between the $\gamma\gamma$ system and
the $W$ in \figre{fig:histo2}. To show the radiation zero most clearly not
only photon isolation but also jet veto cuts are used.
The variable used here is similar to that used for the Tevatron analysis of
the radiation zero in \mref{Baur:1997bn}.
At leading order a dip at zero pseudorapidity can be seen which
is stronger for photons in the same hemisphere since the exact amplitude
zero only occurs for collinear photons.
The NLO corrections have the effect of almost filling the dip, similar to
what has been observed for $W\gamma$ production at the
LHC~\cite{Baur:1993ir}. This will make it very challenging to detect this
feature of the amplitude at the LHC.


\section{Conclusions}

We have calculated the NLO QCD corrections to $\wyy$ production at the LHC.
This process can be used to constrain the anomalous quartic $\wwyy$ gauge
couplings and is therefore an important testing ground of the Standard
Model. We have also included the single fragmentation contribution to this
process at leading order. The fragmentation contribution can however be
effectively suppressed using photon isolation cuts.

We find large radiative corrections of about +200\% which reduce to about
+35\% when using an additional jet veto.  The size of the corrections is
similar to what has been found in $W\gamma$ and $Z\gamma$ production.  As
expected the total scale dependence is increased at NLO while the pure
factorization scale dependence is stabilized. The corrections induce strong
shape distortions in several key distributions. They also tend to fill the
dip caused by the radiation zero of the amplitude making it very hard to
experimentally verify this feature at the LHC. The calculation has been
implemented in a flexible Monte Carlo code which allows to study any
distribution using arbitrary cuts.

\acknowledgments

This work is supported in part by the National Science Foundation under
grant no. NSF-PHY-0547564 and NSF-PHY-0757691.


\bibliographystyle{h-physrev}
\bibliography{main}

\end{document}